\documentclass{JHEP3}

\usepackage{amsmath,amssymb,graphics}

\usepackage{epsfig,multicol}

\title{Quantization of Black Hole Entropy from Quasinormal Modes  }
\author{Shao-Wen Wei,
        Ran Li,
        Yu-Xiao Liu\footnote{Corresponding author.},
        Ji-Rong Ren\\
    Institute of Theoretical Physics, Lanzhou University,
           Lanzhou 730000, P. R. China\\
    E-mail: \email{weishaow06@lzu.cn},
            \email{liran05@lzu.cn},
            \email{liuyx@lzu.edu.cn}}

\abstract{In Phys. Rev. D \textbf{78} (2008) 104018
[arXiv:0807.1481], the conclusion that ``entropy eigenvalues of GB
black hole are discrete and equally spaced, but the area spacing is
not equidistant'' was firstly presented by Kothawala, Padmanabhan
and Sarkar. In this paper, using the new physical interpretation of
quasinormal modes proposed by Maggiore, we calculate the quantum
spectra of entropy for various types of non-rotating black holes
with no charge. The spectrum is obtained by imposing Bohr-Sommerfeld
quantization condition to the adiabatic invariant quantity. We
conjecture that the spacing of entropy spectrum is equidistant and
is independent of the dimension of spacetime. However, the spacing
of area spectrum depends on gravity theory. In Einstein's gravity,
it is equally spaced, otherwise it is non-equidistant. This
conjecture agrees with the result of Kothawala, Padmanabhan and
Sarkar.}


\begin{document}

\section{Introduction}
\label{secIntroduction}

Bekenstein \cite{Bekenstein1} conjectured that, in a quantum
gravity theory, the black hole area should be represented by a
quantum operator with a discrete spectrum of eigenvalues. By
supposing that the black hole horizon area is an adiabatic
invariant, he showed that the area spectrum of black hole is
equidistant and is of the form
\begin{equation}
 \mathcal{A}_{n}=\epsilon \hbar \cdot n,\;\quad  n=0,1,2,...\;.
\end{equation}
This rejuvenates the interest of investigation for quantization of
black hole area
\cite{Hodprl1998,Dreyerprl2003,Polychronakosprd2004,Setarecqg2004,
Setareprd2004,Lepeplb2005,Vagenasjhep2008,Medvedcqg2008,Kothawalaprd2008,Tanaka2008}.
The spacing $\epsilon \hbar$ of area spectrum has been somewhat
controversial. Hod suggested that $\epsilon$ can be determined by
utilizing the quasinormal mode frequencies of an oscillating black
hole \cite{Hodprl1998}. Kunstatter pointed that, for a system with
energy $E$ and vibrational frequency $\Delta\omega(E)$, the ratio
$\frac{E}{\Delta\omega(E)}$ is an adiabatic invariant. He replaced
$E$ with $M$ and identified $\Delta \omega(E)$ as the most
appropriate choice for the frequency. So, by way of
Bohr-sommerfeld quantization condition, one could derive the
spectrum form $\mathcal{A}_{n}=4 \hbar \ln 3 \cdot n$, the same
with Hod's result. Recently, Maggiore in \cite{Maggioreprl2008}
argued that, in high damping limit, the proper frequency of the
equivalent harmonic oscillator, i.e., the quasinormal mode
frequencies $\omega(E)$, should be of the form
$\omega(E)=\sqrt{|\omega_{R}|^{2}+|\omega_{I}|^{2}}$ rather than
the real part $\omega_{R}$. It is clear that when
$\omega_{I}\rightarrow 0$, one could get $\omega(E)=\omega_{R}$
approximately which was adopted extensively in
\cite{Hodprl1998,Dreyerprl2003,Setarecqg2004}. However, under the
case of large $n$ limit or highly excited quasinormal modes for
which $\omega_{R}\ll \omega_{I}$, the frequency of the harmonic
oscillator becomes $\omega(E)= |\omega_{I}|$.

Motivated by this idea, Medved and Vagenas
\cite{Vagenasjhep2008,Medvedcqg2008} made the choice that the
vibrational frequency
$\Delta\omega(E)=(|\omega_{I}|)_{m}-(|\omega_{I}|)_{m-1}$ and
obtained the area spectrum for kerr black hole. They found there
exists a logarithmic term in the adiabatic invariant, which leads
to the non-equidistant area spectrum. In \cite{Setarecqg2004},
Setare calculated the area spectrum for non-rotating BTZ black
hole, and the spectrum is non-equidistant spaced. This result is
in contrast with the area spectrum of black hole in higher
dimension. For Gauss-Bonnet (GB) gravity, for example a
5-dimensional Gauss-Bonnet black hole, the spectra of area and
entropy were obtained from the quasinormal modes by Kothawala et
al. [10]. They concluded that the entropy spectrum is discrete and
equidistant, but the spacing of area spectrum is not equidistant.
Here, one may ask what do other non-rotating black holes with no
charge behave and whether the spacing of area or entropy spectrum
depends on the dimension of spacetime and gravity theory. In order
to answer these questions, we investigate the area and entropy
spectra of 3-dimensional non-rotating BTZ black hole,
4-dimensional Schwarzschild black hole and 5-dimensional GB black
hole with the choice suggested by Medved and Vagenas that
$\Delta\omega(E)=(|\omega_{I}|)_{m}-(|\omega_{I}|)_{m-1}$. Our
results show that, in Einstein¡¯s gravity theory, the spacings of
area and entropy spectra are discrete and equally spaced. For the
GB black hole, we give a explicit calculation following [10] and
our investigations also support the result of Kothawala et al..
However, when setting the GB coupling constant
$\alpha_{GB}\rightarrow 0$, the spacing of area spectrum becomes
equidistant. In summary, for non-rotating black holes with zero
charge, we conjecture that the spacing of entropy spectrum is
equidistant and independent of the dimension of spacetime and
gravity theory. But the spacing of area spectrum depends on
gravity theory. In Einstein's gravity, it is equally spaced,
otherwise it is non-equidistant.

This paper is organized as follows. In section \ref{Schwarzschild
black hole}, we briefly review the method used in this paper and
obtain the entropy quantization of Schwarzschild black hole. The
discussions for 3-dimensional non-rotating BTZ  black hole and
5-dimensional GB black hole appear in sections \ref{non-rotating
BTZ  black hole} and \ref{Gauss-Bonnet black hole}. Finally, the
paper ends with a brief conclusion.

\section{Entropy quantization of 4-dimensional Schwarzschild black hole}
\label{Schwarzschild black hole}

In this section, we modify the frequency that appears in the
adiabatic invariant. Then through calculating the adiabatic
invariant of Schwarzschild black hole, we manage to obtain the
entropy and area spectrum.

Firstly, we consider 4-dimensional Schwarzschild black hole, which
is charactered by the metric
\begin{equation}
 {\mathrm d}s^{2}=-\left(1-\frac{2 M}{r}\right)dt^{2}
    +\frac{1}{1-\frac{2M}{r}} dr^{2}+r^{2}d\Omega_{2},
\end{equation}
where $M$ is the mass of black hole. The radius of the event
horizon $r_{h}$ is $r_{h}=2M$. The surface gravity
$\kappa=\frac{1}{4M}$. Area and Hawking temperature for this black
hole are given by
\begin{equation}
 \mathcal{A}=4\pi r_{h}^{2}=16\pi M^{2}
\end{equation}
and
\begin{equation}
 T=\frac{\kappa}{2\pi}=\frac{1}{8\pi M},
\end{equation}
respectively.

In Ref. \cite{Hodprl1998}, Hod succeeded in deriving the quantum
of the area spectrum using the Bohr's Correspondence principle
from quasinormal modes. The complex quasinormal modes that
correspond to the perturbation equation of Schwarzschild black
hole are also obtained:
\begin{equation}
 M\omega_{m}=0.0437123-\frac{i}{4}\left(m+\frac{1}{2}\right)
    +{\cal{O}}\left[(m+1)^{-\frac{1}{2}}\right]. \label{schwarzschild}
\end{equation}
Noting that the highly damped ringing frequencies depend only upon
the black hole mass. This feature is consistent with the
interpretation of the highly damped ringing frequencies as
characteristics of the black hole itself in the $m \gg 1$ limit.

Kunstatter proposed that given a system with energy $E$ and
vibrational frequency $\Delta\omega(E)$, a natural adiabatic
invariant quantity is \cite{Kunstatterprl2003}:
\begin{equation}
 I=\int \frac{dE}{\Delta\omega(E)}.\label{adiabatic}
\end{equation}
In the large $n$ limit, the Bohr-Sommerfeld quantization can be
expressed as
\begin{equation}
 I=n\hbar. \label{condition}
\end{equation}
Making the choice $\Delta\omega(E)\approx\frac{\ln 3}{8\pi M}$, one
can obtain the area spectrum of a Schwarzschild black hole
\begin{equation}
 \mathcal{A}_{n}=4\ell_{p}^{2}\ln 3\cdot n,
\end{equation}
where $\ell_{p}$ is the Planck length. It can be seen that the area
spectrum is equally spaced with spacing $4\ell_{p}^{2}\ln 3$ and in
agree with the Bekenstein's conjecture.

Recently, Maggiore refined Hod's treatment by arguing that the
physically relevant frequency would actually be
\cite{Maggioreprl2008}
\begin{equation}
 \omega(E)=\sqrt{|\omega_{R}|^{2}+|\omega_{I}|^{2}},
\end{equation}
where $\omega_{R}$ and $\omega_{I}$ are the real and imaginary
parts of the quasinormal mode frequency respectively. When
$\omega_{I}\rightarrow 0$, one could get $\omega(E)=|\omega_{R}|$
approximately. However, under the case of large $m$ or highly
excited quasinormal modes for which $\omega_{R}\ll \omega_{I}$,
the frequency of the harmonic oscillator becomes $\omega(E)=
|\omega_{I}|$. With this supposition, Vagenas and Medved obtained
the area spectrum of Kerr black hole. They calculated the
adiabatic invariant quantity $I$ and area spectrum, which are
given by \cite{Vagenasjhep2008,Medvedcqg2008}
\begin{eqnarray}
 &&I =\frac{\mathcal{A}}{4\pi}-2M^{2}
      \log \big(\frac{\mathcal{A}}{8\pi}\big), \label{logarithimc}\\
 &&\mathcal{A}_{n} + {\cal O}(J_n^4) = 4 \pi \ell_{Pl}^{2}\cdot n.
\end{eqnarray}
This area spectrum is non-equidistant and depends on the angular
momentum of Kerr black hole. It is shown that, when the angular
momentum $J$ is small, the area spectrum is equidistant.

Here, we want to ask what will happen for Schwarzschild black hole
under this supposition. Now, let us turn back to the Schwarzschild
black hole. The adiabatic invariant quantity $I$ for Schwarzschild
black hole is
\begin{equation}
 I=\int \frac{dM}{\Delta\omega}. \label{schwarzschild I}
\end{equation}
From the quasinormal modes (\ref{schwarzschild}), vibrational
frequency $\Delta\omega$ can be obtained
\begin{equation}
 \Delta\omega=(|\omega_{I}|)_{m}-(|\omega_{I}|)_{m-1}=\frac{1}{4M}.
\end{equation}
Substituting it into (\ref{schwarzschild I}) and using the
Bohr-Sommerfeld quantization condition (\ref{condition}), one can
obtain the area spectrum for Schwarzschild black hole
\begin{equation}
 \mathcal{A}_{n}=8 \pi \ell_{Pl}^{2}\cdot n,
\end{equation}
which precisely coincides with Bekenstein's result. The entropy
spectrum can be obtained from the relation $S=\frac{A}{4}$. The
entropy spectrum is given by
\begin{equation}
 S_{n}=2 \pi \ell_{Pl}^{2}\cdot n,
\end{equation}
which is an equidistant spectrum.

\section{Entropy quantization of (2+1)-dimensional non-rotating BTZ  black hole}
\label{non-rotating BTZ  black hole}

Now, in this section, we would like to apply the method to deal with
a (2+1)-dimensional non-rotating BTZ black hole. The line element
for non-rotating BTZ black hole is
\begin{equation}
  ds^2 =-\left(-M+ \frac{r^2}{l^2} \right )dt^2
           +\frac{dr^2}{\left(-M+\frac{r^2}{l^2}\right)}
           +r^{2} d\theta ^2,  \label{metric}
\end{equation}
where $M$ is the Arnowitt-Deser-Misner (ADM) mass and cosmological
constant is given by $\Lambda=\frac{1}{\ell^{2}}$. The event horizon
locates at
\begin{equation}
 r_{h}=\sqrt{\frac{M}{\Lambda}},
\end{equation}
and its area is given by
\begin{equation}
 \mathcal{A}=2\pi r_{h}=2\pi\sqrt{\frac{M}{\Lambda}}. \label{BTZarea}
\end{equation}
The quasinormal frequency for non-rotating BTZ black hole has been
obtained by Cardoso and Lemos in \cite{Cardosoprd2001}
\begin{equation}
 \omega=\pm m-2iM^{1/2}(m+1), \quad m=0,1,2,...~.
\end{equation}
The adiabatic invariant quantity $I$ for non-rotating BTZ black
hole is
\begin{equation}
 I=\int \frac{dM}{\Delta\omega}. \label{BTZ}
\end{equation}
At large $m$, the vibrational frequency $\Delta\omega$ is
\begin{equation}
 \Delta\omega=(|\omega_{I}|)_{m}-(|\omega_{I}|)_{m-1}=2\sqrt{M}.
\end{equation}
Substituting $\Delta\omega$ into (\ref{BTZ}), we obtain the
adiabatic invariant quantity
\begin{equation}
 I=\sqrt{M}.
\end{equation}
Using Bohr-Sommerfeld quantization condition (\ref{condition}), we
have
\begin{equation}
\sqrt{M}=n \hbar.
\end{equation}
Recalling the area from (\ref{BTZarea}), we derive the area spectrum
of this black hole
\begin{equation}
 \mathcal{A}_{n}=2\pi n \hbar\sqrt{\frac{1}{\Lambda}}. \label{BTZresult}
\end{equation}
It is clear that the cosmological constant $\Lambda$ appears in
the area spectrum (\ref{BTZresult}). We also note that this area
spectrum is equally spaced. In Ref. \cite{Setarecqg2004}, the
result is $\mathcal{A}_{n}=2\pi \sqrt{\frac{nm \hbar}{\Lambda}}$.
Although this spectrum is quantized, it is not equally spaced.
This conflicts with the conjecture of Bekenstein. In Ref.
\cite{Maggioreprl2008} Maggiore argues that, under high damped
modes, it is not accurate to make $\omega=\omega_{R}$. For this
case, one needs take
$\omega=(\omega_{R}^{2}+\omega_{I}^{2})^{\frac{1}{2}}$. Through
taking the vibrational frequency
$\Delta\omega=(|\omega_{I}|)_{m}-(|\omega_{I}|)_{m-1}$, we get the
area spectrum (\ref{BTZresult}), which has equidistant $\Delta
\mathcal{A}=2\pi \hbar\sqrt{\frac{1}{\Lambda}}$. Entropy spectrum
of this black hole is
\begin{equation}
 S_{n}=\frac{1}{2}\pi \hbar\sqrt{\frac{1}{\Lambda}}\cdot n ,
\end{equation}
with the spacing
\begin{equation}
 \Delta S=S_{n+1}-S_{n}=\frac{1}{2}\pi \hbar\sqrt{\frac{1}{\Lambda}}.
\end{equation}
It is clear that this entropy is equidistant and the spacing
depends on the cosmological constant $\Lambda$.

\section{Entropy quantization of 5-dimensional Gauss-Bonnet black hole}
\label{Gauss-Bonnet black hole}

In this section, following the calculation [10], we recalculate
and obtain a explicit form of entropy and area quantization of
black hole in GB gravity theory. The (4+1) dimensional static,
spherically symmetric black hole solution in this theory is of the
form
\begin{eqnarray}
{\mathrm d}s^2 &=& - f(r) {\mathrm d}t^2 + f(r)^{-1}{\mathrm d}r^2 +
r^2 {\mathrm d}\Omega_{3},
\end{eqnarray}
where the metric function is
\begin{eqnarray}
 f(r) &=& 1 + \frac{r^2}{ 2 {\alpha} } \left[ 1 - \left( 1 + \frac{4
       ~ \alpha ~ \varpi}{r^{4}} \right)^{1/2} \right].
\end{eqnarray}
Here, $\alpha=2\alpha_{GB}$ and $\varpi$ is related to the ADM mass
$M$ by the relationship $\varpi = \frac{ 16 \pi }{ 3 \Sigma_{3} } ~
M$, where $\Sigma_{3}$ is the volume of unit 3 sphere. The event
horizon is located at $r=r_{h}$, and $r_{h}$ satisfies
\begin{equation}
 r_{h}^{2} + \alpha  - \varpi=0.
\end{equation}
For the horizon to exist at all, one must have $r_h^2 + 2 \alpha
\geq 0$.

Area and Hawking temperature for this black hole are given,
respectively, by
\begin{equation}
 \mathcal{A}=2 \pi^{2} (-\alpha+\frac{8M}{3\pi})^{3/2},
 \label{GBqarea}
\end{equation}
and
\begin{equation}
 T = \frac{1}{2 \pi} \frac{ r_{h} }{(r_{h}^{2} + 2 \alpha)}.
\end{equation}
The highly damped quasinormal modes for 5-dimensional GB black
hole (when $\omega_{I} \gg \omega_{R}$) have been worked out
\cite{Daghighcqg2007}
\begin{equation}
\omega(m) \underset{m \rightarrow \infty}{\longrightarrow} T
\ln{Q} + i (2 \pi T) m. \label{GBspectrum}
\end{equation}
The imaginary part can be understood  in terms of a scattering
matrix formalism; see e.g., \cite{Padmanabhancqg2004}. We identify
the relevant frequency as $\omega=|\omega_{I}|$. The adiabatic
invariant quantity $I$ for this black hole is
\begin{equation}
 I=\int \frac{dM}{\Delta\omega}. \label{GB}
\end{equation}
From the quasinormal modes (\ref{GBspectrum}), vibrational frequency
$\Delta\omega$ can be evaluated as
\begin{equation}
 \Delta\omega=(|\omega_{I}|)_{m}-(|\omega_{I}|)_{m-1}=2\pi T.
\end{equation}
Using the expression of event horizon area for GB black hole, the
adiabatic invariant is rewritten as
\begin{equation}
 I=\frac{8 M+15\pi \alpha}{12}\sqrt{\frac{8M}{3}- \alpha}.
\end{equation}
With Bohr-Sommerfeld quantization condition (\ref{condition}) and
the area (\ref{GBqarea}), we obtain the area spectrum
\begin{equation}
 \mathcal{A}_{n}+6\alpha(4\pi^{4}\mathcal{A}_{n})^{\frac{1}{3}}=8\pi \hbar\cdot n. \label{GBarea1}
\end{equation}
The spacing of this spectrum is
\begin{equation}
 \Delta \mathcal{A}= \mathcal{A}_{n+1}-\mathcal{A}_{n}=8\pi
                       \hbar+g(\mathcal{A}_{n,n+1}), \label{GBarea}
\end{equation}
where, the function $g(\mathcal{A}_{n,n+1})$ is
\begin{equation}
 g(\mathcal{A}_{n,n+1})=6\alpha(4\pi^{4})^{\frac{1}{3}}(\mathcal{A}_{n}^{\frac{1}{3}}-\mathcal{A}_{n+1}^{\frac{1}{3}}).
\end{equation}
The function $g(\mathcal{A}_{n,n+1})$ is a correction term to the
spacing of the spectrum, and leads to a non-equidistant area
spectrum.

In Einstein's gravity, entropy of the horizon is proportional to
its area. Equidistance of the area spectrum implies that the
entropy spectrum is also equidistant. However, when one considers
the natural generalization of Einstein gravity by including higher
derivative correction terms like the GB term to the original
Einstein-Hilbert action, the trivial relationship $S=\frac{A}{4}$
between horizon area and associated entropy does not hold anymore.
The relationship is now
\begin{equation}
 S = \frac{\mathcal{A}}{4} \left[ 1
    + 6 \alpha \left(\frac{\mathcal{A}}{\Sigma_{3}} \right)^{-2/3}
    \right]. \label{GBentropy}
\end{equation}
Substituting (\ref{GBarea1}) into (\ref{GBentropy}), we obtain the
entropy spectrum of GB black hole
\begin{equation}
 S_{n}=2\pi\hbar\cdot n.
\end{equation}
It is clear that this entropy spectrum is equally spaced with
$\Delta S=2\pi\hbar$. The results are in agreement with that of
Kothawala et al. [10]. They firstly pointed out the notions that,
for GB gravity, the entropy eigenvalues are discrete and equally
spaced, but the area spacing is not equidistant and quantum of
entropy is more appropriate than the quantum of area. One can see
that, the area spectra for 3-dimensional non-rotating BTZ black
and 4-dimensional Schwarzschild black hole are all discrete and
equally spaced. However, these characters partially hold for
5-dimensional GB black hole. It seems that, whether the area
spectrum is equidistant does not depend on the dimension of
spacetime, but depend on gravity theory. It is easy to see that,
if we set the GB coupling constant $\alpha_{GB}$ to 0, the
equation (\ref{GBentropy}) shows a linear relationship between
entropy and area. Then equidistant entropy spectrum implies the
equidistant area spectrum. The equidistant area spectrum also can
be seen from function $g(\mathcal{A}_{n,n+1})=0$ when GB coupling
constant $\alpha_{GB}\rightarrow 0$.

\section{Conclusion}
\label{secConclusion}

In summary, by modifying the frequency that appears in the
adiabatic invariant of black hole and using the Bohr-Sommerfeld
quantization under large $n$ limit, we investigate the entropy and
area spectra of Schwarzschild black hole, (2+1)-dimensional
non-rotating BTZ black hole and 5-dimensional GB black hole,
respectively. All these results imply that the entropy for
different types of black holes can be quantized and equally
spaced. The area can also be quantized, but the spacing depends on
the gravity theory. In Einstein's gravity, area spectrum is
equally spaced, but in GB gravity, the spacing of the area
spectrum is non-equidistant. Furthermore, for the non-rotating BTZ
black hole, the spacings of both entropy and area spectra depend
on the cosmological constant $\Lambda$. So this result may imply
some intrinsic characteristics of the non-rotating BTZ black hole.
It is worth to point out that the result we obtained is only for
non-rotating black holes with no charge. For more general cases,
the investigation should be carried out in our further work. We
should keep in mind that, all our's calculations are semiclassical
and based on Bohr-Sommerfeld quantization condition and the
quasinormal modes. These results are just conjectural and the
underlying relationship between entropy quantization and
quasinormal modes should be given more attention.

\section*{\textbf{Note}}

The result ``entropy eigenvalues of GB black hole are discrete and
equally spaced, but the area spacing is not equidistant" in
Section 4 was firstly put forward by Dawood Kothawala et al. in
Ref. [10]. We also thank Dawood Kothawala et al. for useful
comment.

\section*{Acknowledgements}

This work was supported by the National Natural Science Foundation
of China (No. 10705013), the Doctor Education Fund of Educational
Department of China (No. 20070730055) and the Fundamental Research
Fund for Physics and Mathematics of Lanzhou University (No.
Lzu07002).

\end{document}